\documentclass[journal=jacsat,manuscript=article]{achemso}

\usepackage{chemformula} 
\usepackage[T1]{fontenc} 

\author{A. Saeidi}
\email{ali.saeidi@epfl.ch}
\author{F. Jazaeri}
\author{I. Stolichnov}
\author{Christian C. Enz}
\author{Adrian M. Ionescu}
\affiliation[Ecole Polytechnique Federale de Lausanne (EPFL)]{Ecole Polytechnique Federale de Lausanne (EPFL), Lausanne, Switzerland}

\title[An \textsf{achemso} demo]
  {Negative Capacitance as Digital and Analog Performance Booster for Complementary MOS Transistors}

\abbreviations{IR,NMR,UV}

\begin{document}


\begin{abstract}
	Boltzmann tyranny poses a fundamental limit to lowering the energy dissipation of conventional MOS devices, a minimum increase of the gate voltage, i.e. 60 mV, is required for a 10-fold increase in drain-to-source current at 300 K. Negative Capacitance (NC) in ferroelectric materials is proposed in order to address this physical limitation of CMOS technology. A polarization destabilization in ferroelectrics causes an effective negative permittivity, resulting in a differential voltage amplification and a reduced subthreshold swing when integrated into the gate stack of a transistor. Recent demonstrations of negative capacitance concerned mainly n-type MOSFETs and their subthreshold slope. An effective technology booster should be capable of improving the performance of both n- and p-type transistors. In this work, we report a significant enhancement in both digital (subthreshold swing, on-current over off-current ratio, and overdrive) and analog (transconductance and current efficiency factor) FoM of commercial 28nm CMOS process by exploiting a PZT capacitor as the negative capacitance booster. Accordingly, a sub-thermal swing down to 10 mV/decade together with an enhanced current efficiency factor up to 10$^5$ V$^{-1}$ is obtained in both n- and p-type MOSFETs at room temperature. The overdrive voltage is enhanced up to 0.45 V, leading to a supply voltage reduction of 50\%. 
\end{abstract}

\section{}

Complementary Metal-Oxide-Semiconductor (CMOS) scaling will be eventually limited by the inability to remove the heat generated in the switching process \cite{takagi2008carrier}. The origin of this issue can be traced back to the operation principle of the silicon CMOS devices governs by the non-scalability of thermal voltage (Boltzmann's tyranny). This results in preventing these devices to achieve a sub-60 mV/decade subthreshold slope (SS) at room temperature. The SS of a MOSFET is obtained by
\begin{equation}
SS=\frac{\partial V_g}{\partial (logI_d)}=\frac{\partial V_g}{\partial \psi _s}\times\frac{\partial \psi _s}{\partial (logI_d)},
\end{equation}
where $\psi _s$ corresponds to the surface potential of the silicon channel. In a conventional MOSFET, the lower limit of the second term in RHS of (1) is  $(k_BT/q)Ln(10)$ and cannot be any lower than 60 mV/decade at 300 K. Since V$_g$ is linked to $\psi _s$ through a capacitive voltage divider, the first term that is known as the body factor, m, is obtained as
\begin{equation}
\frac{\partial V_g}{\partial \psi _s}=1+\frac{C_s}{C_{MOS}},
\end{equation}
exceeds one, thus limits the SS to 60 mV/decade at T=300 K \cite{khandelwal2016circuit,ionescu2011ultra}. A sub-thermal swing can be achieved using the proposed negative capacitance (NC) of ferroelectric materials \cite{salahuddin2008can,salahuddin2008use}. Negative capacitance in ferroelectrics arises from the imperfect screening of the spontaneous polarization. Imperfect screening is intrinsic to semiconductor-ferroelectric and metal-ferroelectric interfaces due to their screening lengths. The physical separation of ferroelectric bound charges from the metallic screening charges creates a depolarizing field inside the ferroelectric and destabilizes the polarization \cite{zubko2016negative}. Hence, intentionally destabilizing this polarization causes an effective NC that has been proposed as a way of overcoming the fundamental limitation on the power consumption of MOSFETs \cite{jo2015negative,gao2014room,ionescu2018negative}. The negative capacitance, originating from the dynamics of the stored energy in a phase transition of ferroelectric materials, results in an internal voltage amplification in an MOS device when integrated into the gate stack. The concept of NC can be understood by considering the free energy of ferroelectrics. A ferroelectric material is traditionally modeled using a double well energy landscape. The energy characteristic of a ferroelectric capacitor, depicted in Figure 1a, is calculated by $U_{FE}=\alpha P^2+\beta P^4+\gamma P^6+E_{ext}.P$, where $P$ is the polarization, $E_{ext}$ is the externally applied electric field, and $\alpha$, $\beta$, and $\gamma$ are material dependent parameters \cite{salahuddin2008use}. In equilibrium, the ferroelectric resides in one of the wells, providing spontaneous polarization. The capacitance of a ferroelectric material can be determined by 
\begin{equation}
C_{FE}=\left[\frac{d^2U_{FE}}{dQ_{FE}^2}\right]^{-1},
\end{equation}
which is positive at the wells considering the curvature of U$_{FE}$ vs. Q$_{FE}$ characteristic (Figure 1a). Nevertheless, the curvature is negative around the origin (Q$_{FE}$ = 0). More specifically, a ferroelectric material shows an effective NC while switching from one stable polarization state to the other one \cite{khan2015negative}. It should be remarked that NC refers to negative differential capacitance due to the small signal concept of the capacitance and relation between C$_{FE}$ and U$_{FE}$ (equation 3). The NC has been proven elusive for ferroelectrics in isolation and cannot be observed in experiments, exhibiting hysteretic jumps in the polarization (Figure 1b). However, it has been confirmed that if the ferroelectric placed in-series with a positive capacitor of proper value, the NC segment can be stabilized \cite{appleby2014experimental,yeung2013low}. This NC region can be modeled by the state-of-the-art approach for modeling the dynamics of ferroelectric capacitors relying on Landau-Khalatnikov (LK) equation, $\rho (dP/dt)+\nabla _pU_{FE} = 0$. Figure 1b compares the experimentally measured polarization vs. electric field of a PZT capacitor with the fitting result of the LK equation.

\begin{figure}
	\begin{centering}
		\includegraphics[width=10cm,height=4.3cm]{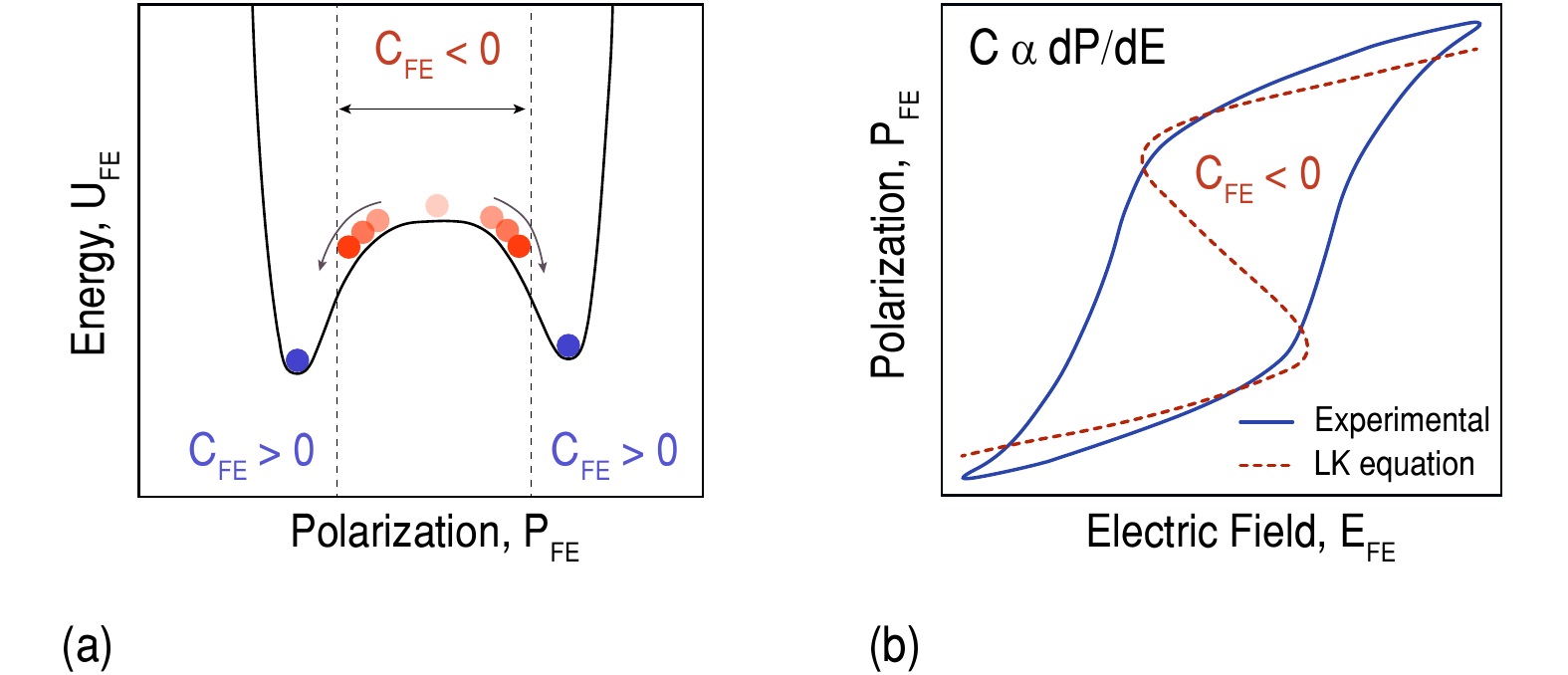}
		\par\end{centering}
	\protect\caption{Free energy landscape and polarization characteristic of a ferroelectric. (a) Energy density function of a ferroelectric capacitor in equilibrium, showing an effective NC while switching from one stable polarization state to the other one. (b) Measured polarization vs. electric field for a PZT film (experimental) and the fitting results of LK equation (dashed curve).}
\end{figure}

A ferroelectric capacitor interconnecting with the gate stack of a MOS transistor creates a series connection between C$_{FE}$ and C$_{MOS}$ (Figure 2a). The ferroelectric capacitor can increase the total capacitance of the gate ($C^{-1}_{total}=C^{-1}_{FE}+C^{-1}_{MOS}$) while it is stabilized in the NC region \cite{jain2014stability,khan2011ferroelectric}. Specifically, the series structure brings an abrupt increase in the differential charge in the internal node (V$_{int}$) by changing the gate voltage, thus providing a step-up voltage transformer \cite{rusu2012analytical,saeidi2017negative2}. The internal gain of NC can be defined as $\beta=\partial V_{int}/\partial V_g=C_{FE}/(C_{FE}+C_{MOS})$. Accordingly, an NC booster can provide an internal voltage amplification ($\beta>1$) which results in a body factor reduction, i.e. $1/ \beta$, leading to the improvement of both analog and digital parameters of the transistor:
\begin{equation}
SS=\left(\frac{\partial logI_{d}}{\partial V_{g}}\right)^{-1}=\frac{\partial V_{int}}{\partial logI_{d}}\times\frac{\partial V_{g}}{\partial V_{int}}=\frac{1}{\beta}\times SS_{ref},
\end{equation} 
\begin{equation}
g_{m}=\frac{\partial I_{d}}{\partial V_{g}}=\frac{\partial I_{d}}{\partial V_{int}}\times\frac{\partial V_{int}}{\partial V_{g}}=\beta \times g_{m-ref}.
\end{equation}

\begin{figure}
	\begin{centering}
		\includegraphics[width=10cm,height=4.3cm]{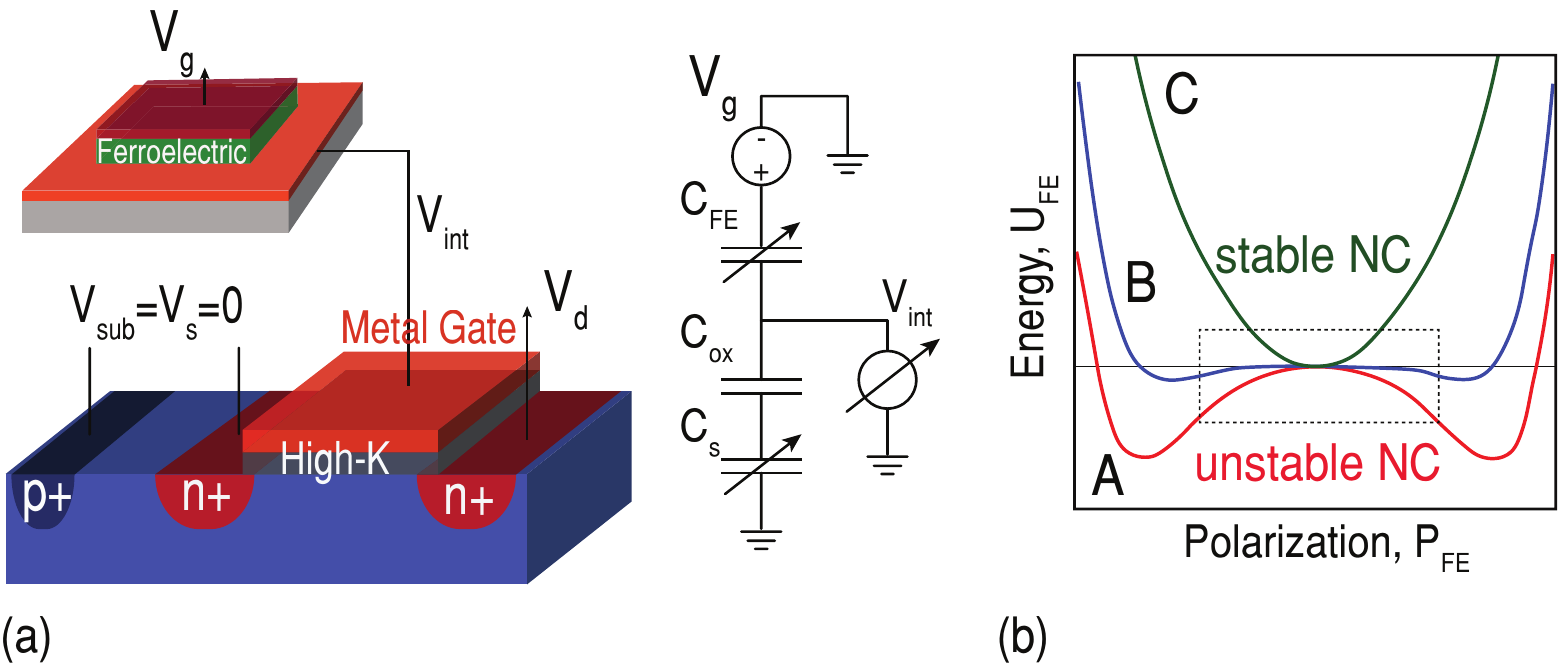}
		\par\end{centering}
	\protect\caption{Negative capacitance MOSFET and stability constraints. (a) Schematic diagram of the experimental configuration of an n-type NC-FET including the capacitance model of the structure. (b) Ferroelectric's NC is unstable by itself (A), but it can be partially (B) or fully stabilized (C) by placing in-series with a positive capacitor. The NC region of ferroelectric energy landscape is shown inside the dotted rectangular box.}
\end{figure} 

In order for NC to occur, the charge line of the baseline transistor is acquired to have an intersection with the negative slope of the polarization \cite{khan2015negative}. Otherwise, the device characteristic shows a hysteresis, corresponding to the coercive fields of the ferroelectric without performance boosting \cite{saeidi2016modeling}. Additionally, to bring about the maximum enhancement in the non-hysteretic operation of an NC-FET, the negative value of C$_{FE}$ should be well-matched with C$_{MOS}$ ($|C_{FE}|$ = $C_{MOS}$ while C$_{total} > $ 0 in the whole range of operation) \cite{saeidi2017negative,saeidi2017negative2}. Generally, considering that the SS can be expressed as $SS=(60 mV/decade).(1+C_{MOS}/C_{FE})$, the transistor transfer characteristic becomes steep as $|C_{FE}|$ gets close to $C_{MOS}$. However, a value of $|C_{FE}|$ too close to $C_{MOS}$ gives rise to the hysteretic behavior due to the instability of NC in the strong inversion regime \cite{saeidi2016double}. Additionally, both C$_{MOS}$ and C$_{FE}$ are voltage-dependent, making it extremely challenging to fully satisfy the matching condition. Therefore, the ferroelectric's NC commonly partially gets stabilized, proposing a trade-off between the hysteretic behavior and the performance-boosting due to the NC effect (Figure 2b). With the validity of NC concept being experimentally established \cite{karda2016switching,yeung2013low,li2015sub,jimenez2010analytic,aziz2016physics,hoffmann2017modeling}, it is now of paramount importance to understand the challenges involved in the design of NC-FETs, so that the steepness and hysteresis of the device characteristic can be optimized in both n- and p-type MOSFETs. In this respect, a polycrystalline PZT capacitor is fabricated for thoroughly understanding the negative capacitance concept. It is then connected to various commercial MOSFETs, fabricated in 28 nm CMOS technology node. The hysteretic behavior of both n- and p-type NC-FETs is tuned imposing the proposed matching condition. Afterward, the impact of NC on the performance of conventional MOSFETs is investigated by measuring and analyzing the internal node voltage. Sub-thermal swing down to 10 mV/decade is observed in n- and p-type hysteretic NC-FETs. The paper reports and discusses the trade-off between the steepness of the subthreshold slope and the hysteresis, degrading the performance by reducing the hysteresis. The strong dependence of the NC effect on the source to drain electric field is also evidenced, reducing the impact by increasing the absolute value of V$_{ds}$. Moreover, it is experimentally validated that a poly-domain ferroelectric capacitor in steady states cannot have more than one stable NC domain at the time, showing a different polarization characteristic from the expected S-shape of a single-domain ferroelectric.

\section{Experimental results and discussion}
As schematically shown in Figure 2a, the experimental results are obtained by connecting an external PZT capacitor to the gate of a MOSFET. This external connection offers a flexibility of testing different series combinations and tuning the hysteretic behavior. High-performance commercial n- and p-type MOSFETs are employed as the baseline transistors. An MIM structure with 45nm of polycrystalline Pb(Zr$_{43}$,Ti$_{57}$)O$_3$ (PZT) is fabricated \cite{kidoh1991ferroelectric,nakamura1994preparation}. High-quality epitaxial ferroelectric layers are commonly considered suitable for NC devices due to the formation of a mono-domain state characterized by a single coercive field \cite{zubko2016negative,khan2016negative,lin2016effects}. However, the typical behavior of poly-domain ferroelectrics can change dramatically by applying a repetitive voltage stress known as the training procedure of ferroelectrics \cite{saeidi2018effect}. This behavior suggests that a poled ferroelectric layer can show a mono-domain like characteristic (see supplementary materials).
\subsection{n-type negative capacitance MOSFETs}

\begin{figure}
	\begin{centering}
		\includegraphics[width=15cm,height=11cm]{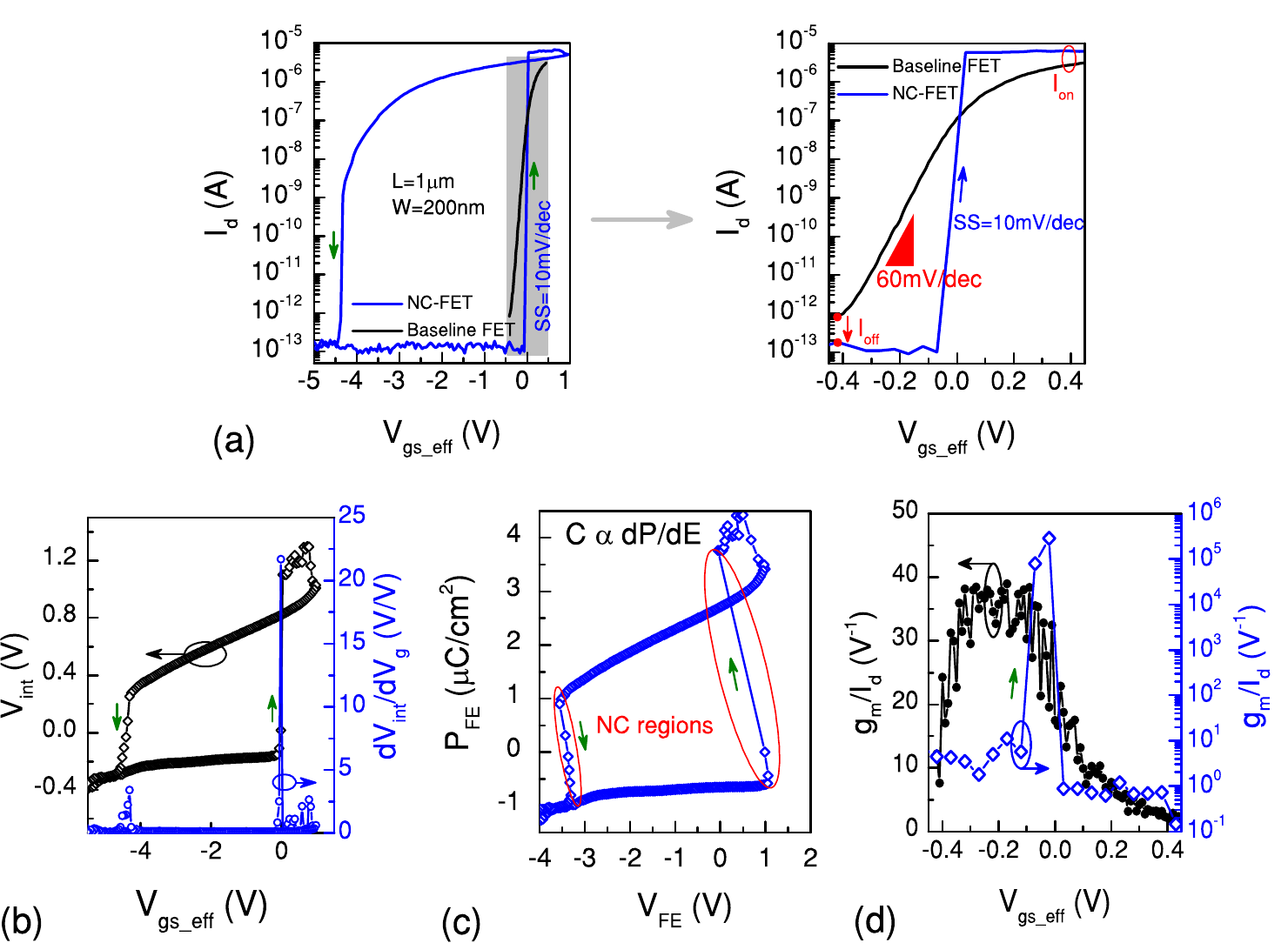}
		\par\end{centering}
	\protect\caption{Hysteretic n-type NC-FET. (a) Transfer characteristic of the device shows a super steep transition of 10 mV/decade together with a hysteresis of 4.5 V (V$_{ds}$ = 100 mV). (b) A remarkable amplification (defined as dV$_{int}$/dV$_g$) up to 20 V/V is achieved in the regions corresponding to the negative slope of the polarization (c). Extracted current efficiency factor of the device represents a significant boosting, up to 10$^5$ V$^{-1}$, in the subthreshold region (d).}
\end{figure}

Figure 3a illustrates the input transfer characteristic of an n-type NC-FET where the gate of the baseline FET (W = 200 nm, L = 1 $\mu$m) is loaded with a PZT capacitor having an area of 30$\times$30 $\mu m^2$. The gate voltage is swept from $-$3 V to $+$3 V and back to $-$3 V while the drain voltage is set to 0.1 V. It should be noted that the curves are plotted with respect to the effective gate voltage (V$_{gs\_eff}$ = V$_{gs}$-V$_{th}$), which makes the results to be directly comparable and removes the effect of the two different threshold voltages. With the aid of an internal electrode, a step-up conversion of the internal voltage is explicitly observed as a result of the ionic movement in PZT. In order to qualitatively determine the voltage gain, dV$_{int}$/dV$_g$ vs. V$_g$ is plotted, representing a significant amplification up to 20 V/V (Figure 3b). This internal voltage increase allows the surface potential to be higher than the gate voltage, leading to a body factor below 1. Therefore, an SS of 10 mV/decade is observed over six orders of magnitude of the drain current. Moreover, the overdrive voltage is improved by a value of 0.45 V. Using the internal electrode and imposing the displacement vector continuity, a negative slope of the polarization is extracted in a certain range of the polarization, corresponding to the subthreshold region where a significant boosting of performance is reached (Figure 3c). It should be noted that due to the charge balance conditions, only a small fraction of the polarization get switched \cite{khan2016negative2}. A remarkable enhancement of the current efficiency factor, g$_m$/I$_d$, with a peak of 10$^5$ V$^{-1}$ is demonstrated when the device is operating in the weak-inversion regime (Figure 3d). A significant improvement is achieved in both digital and analog FoM of the baseline MOSFET. The measured input transfer characteristic of the NC-FET shows a hysteresis of 4.5 V caused by the poor matching of the ferroelectric NC and MOS capacitance \cite{saeidi2017negative,saeidi2016double}.

The undesirable hysteretic operation of NC-FETs can be alleviated with a better matching of the ferroelectric and MOS capacitances which ensures the $C_{total}>0$ stability condition in a wide range of the applied gate voltage \cite{jo2016negative}. Considering $C_{total}^{-1}=C_{FE}^{-1}+C_{ox}^{-1}+C_{si}^{-1}$, where $C_{ox}$ and $C_{si}$ correspond to the gate linear dielectric and silicon capacitances, the stability condition can be written as
\begin{equation}
\left(\frac{S_{gate}}{S_{FE}}\right)<\frac{5\gamma}{4(3\beta ^2-5\alpha \gamma)}\left(\frac{1}{d_{FE}}\right)\left[\frac{d_{ox}}{\epsilon _{SiO_2}}+\frac{d_{si}}{\epsilon _{si}}\right].
\end{equation}

\begin{figure}
	\begin{centering}
		\includegraphics[width=15cm,height=11cm]{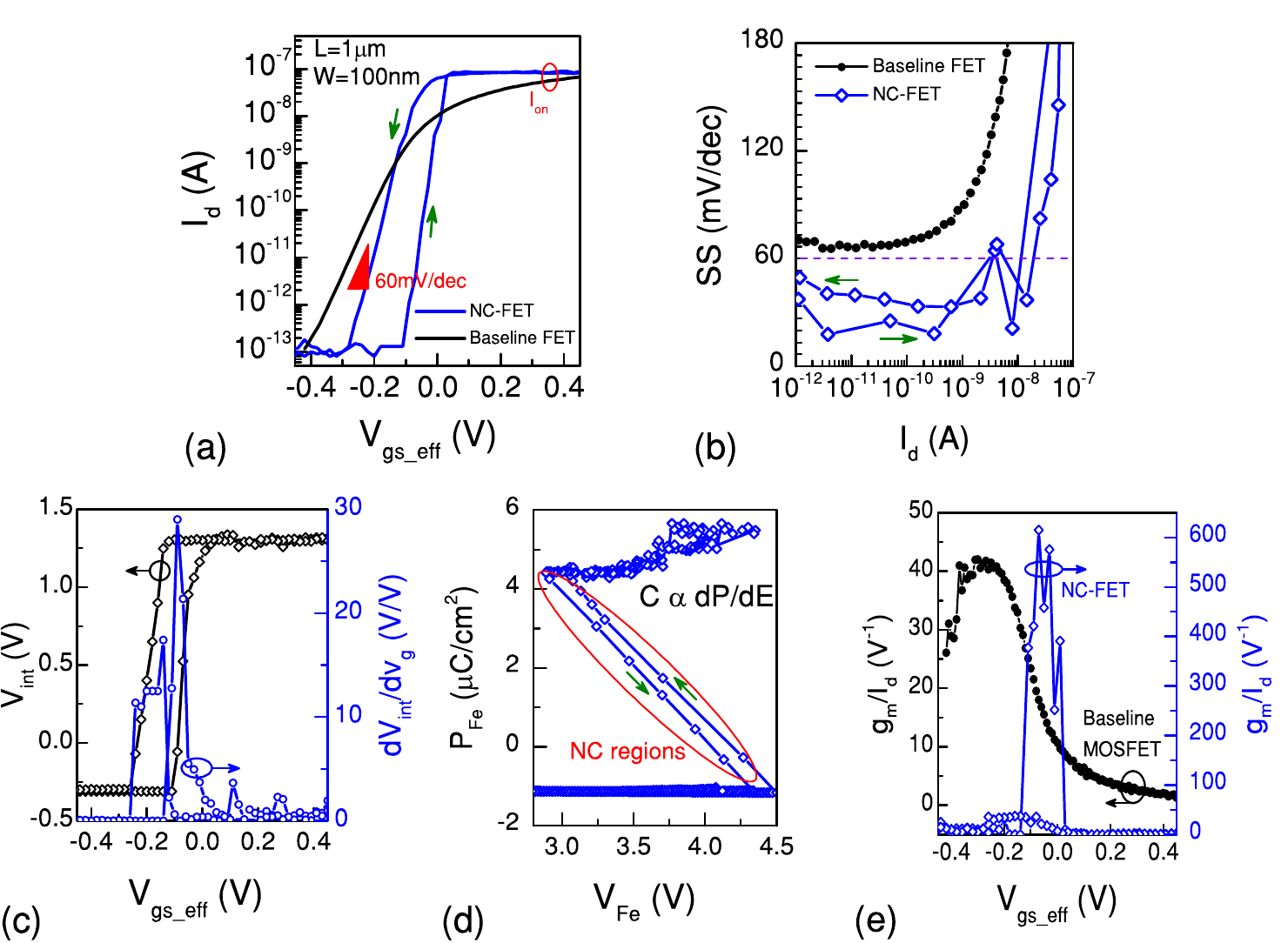}
		\par\end{centering}
	\protect\caption{n-type NC-FET with a reduced hysteresis. (a) Performance of an n-type NC-FET with a small hysteresis of 150 mV and a swing below 30 mV/decade while V$_{ds}$ is set to 100 mV (b). A steep \textit{off}-to-\textit{on} transition is realized in both positive and negative going branches of the drain current. (c) Internal voltage measurement shows a voltage gain of up to 10 V/V. (d) The extracted P-E curve of the ferroelectric shows a clear S-shape in a wide range of operation with a negligible hysteresis. (e) g$_m$/I$_d$ is also boosted and reached a factor of 600 V$^{-1}$.}
\end{figure}

In equation (6), d, S, and $\epsilon$ are the thickness, area, and the permittivity of the corresponding layer, respectively. Hence, in consideration of (6), another NC-FET with a different baseline FET (W = 100 nm, L = 1 $\mu$m) and a PZT capacitor of the same thickness and an area of 20$\times$20 $\mu m^2$ with a better matching of capacitances and a smaller hysteresis is depicted in Figure 4a. A reduced hysteresis of 150 mV is demonstrated while the transistor is operating at a constant drain voltage i.e. 0.1 V. An SS below 30 mV/decade at 300 K is reliably achieved in both positive and negative going branches of the input transfer characteristic (see Figure 4b) where the transistor charge line and the negative slope of the ferroelectric polarization have an intersection \cite{rusu2016condition}. As a result of an average swing well below the thermal limit of MOSFETs, the effective gate voltage can be reduced by 50\%, maintaining the same level of the output current. Figure 4c depicts the internal voltage and internal gain plots (V$_{int}$ vs V$_g$ and dV$_{int}$/dV$_g$ vs V$_g$). Figure 4d depicts the extracted polarization characteristic of the series connected PZT capacitor, showing an effective NC similar to the ideal expectation of NC by LK equation. The current efficiency factor is also enhanced and reached a maximum value of about 600 V$^{-1}$ (Figure 4e).

\subsection{p-type negative capacitance MOSFETs}
The impact of the same NC booster on p-type commercial MOSFETs is also reported and discussed. The drain-to-source voltage was set at $-$0.9 V in all measurements performed in this part, otherwise mentioned. Figure 5 depicts the performance improvement that is obtained in a p-type NC-FET (W = 1 $\mu$m, L = 90 nm) using a PZT capacitor (40$\times$40 $\mu m^2$) as an NC booster. The gate voltage swept from $+$3 V to $-$3 V and returns back to the initial bias by reverse sweep. Using the NC booster, similar to n-type NC-FETs, the internal voltage is enhanced and reached values greater than the applied gate voltage, so that a steep \textit{off}-to-\textit{on} transition of 10 mV/decade is realized over at least 4 orders of magnitude of the drain current (Figure 5a). The NC condition is fulfilled in both forward and reverse sweeps so that a sub-thermal swing is demonstrated in both branches \cite{rusu2016condition}. Due to the poor matching of capacitances, a large hysteresis of 3.5 V is observed. Analyzing the internal electrode voltage (Figure 5b) shows a considerable internal voltage amplification in the regions where the ferroelectric capacitor provides a clear S-shape negative slope of the polarization (Figure 5c). The g$_m$/I$_d$ FoM is also significantly enhanced, reaching a peak of 10$^5$ V$^{-1}$ (Figure 5d).

\begin{figure}
	\begin{centering}
		\includegraphics[width=15cm,height=11cm]{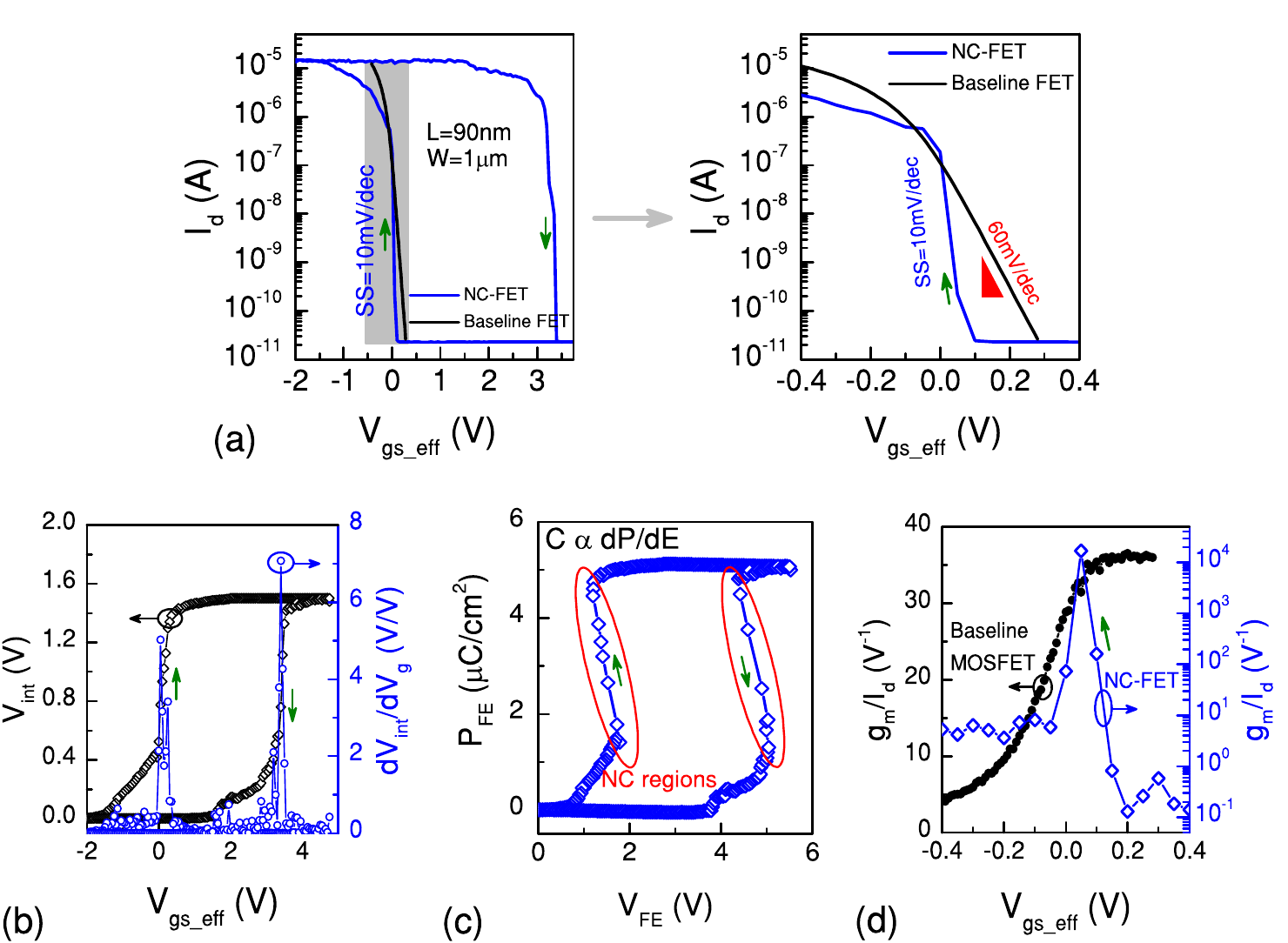}
		\par\end{centering}
	\protect\caption{Hysteretic p-type NC-FET. (a) Transfer characteristic of a p-type NC-FET with a large hysteresis of 3 V ($|V_{ds}|$ = 900 mV) and a swing of 15 mV/dec over five decades of current. (b) An internal voltage gain greater than one is measured in both positive and negative going branches (c). Current efficiency factor is also enhanced, reaching a factor of 10$^5$ V$^{-1}$.}
\end{figure}

\begin{figure}
	\begin{centering}
		\includegraphics[width=15cm,height=11cm]{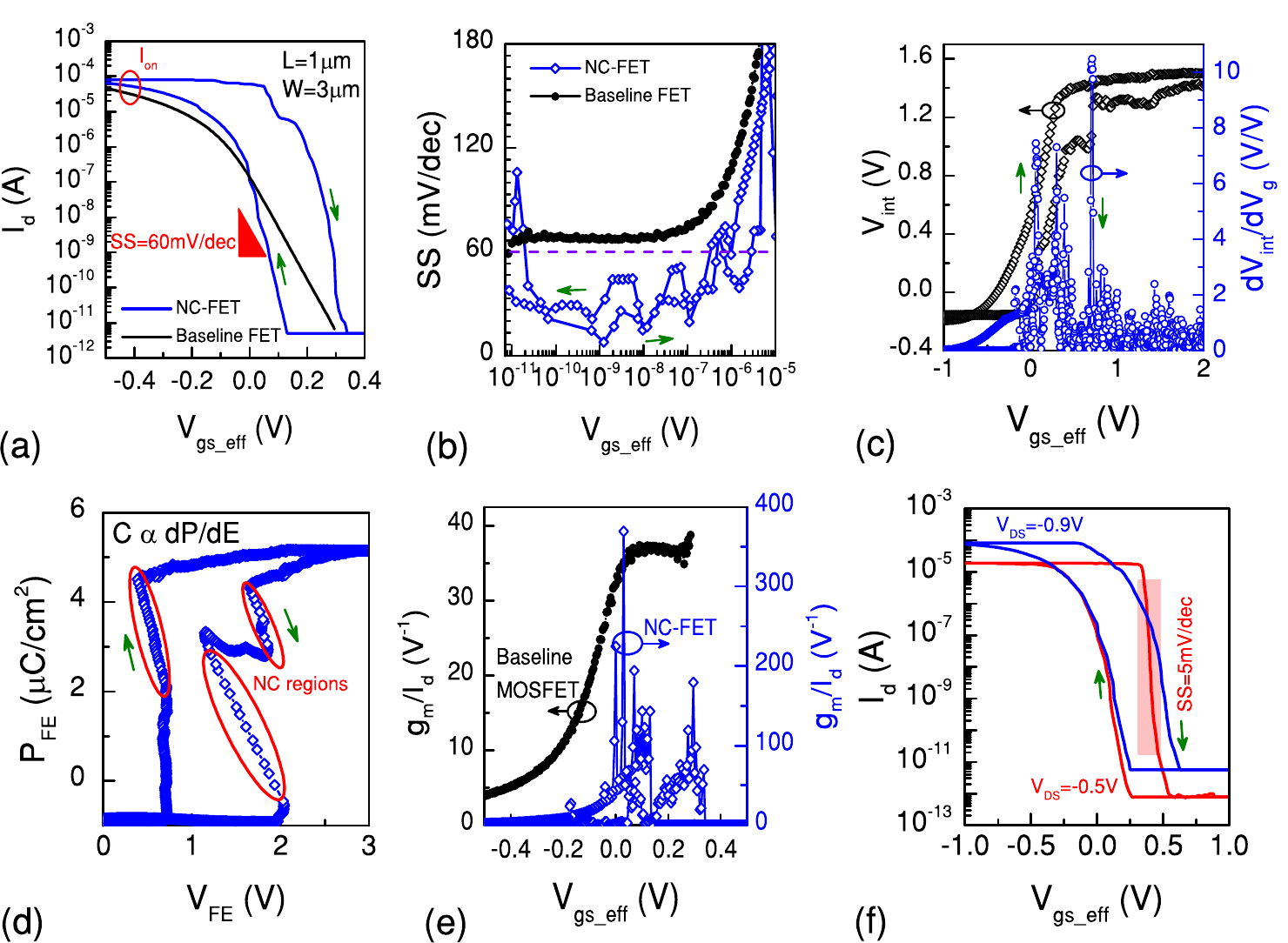}
		\par\end{centering}
	\protect\caption{p-type NC-FET with a reduced hysteresis. (a) Input transfer characteristic of an NC-FET with a small hysteresis of 200 mV at $|V_{ds}|$ = 900 mV. (b) A sub-thermal swing well below 60 mV/dec is obtained. (c) Measurement of the internal node shows a significant voltage gain, having a peak of 10 V/V. (d) Polarization characteristic of the ferroelectric capacitor shows an effective NC in both branches. Two discrete NC regions are observable in the reverse sweep of the gate voltage due to the polycrystallinity of the ferroelectric film. (e) g$_m$/I$_d$ is considerably enhanced and reached a value of 400 V$^{-1}$. (f) shows the impact of the drain-to-source electric field on the steepness and hysteresis of the NC-FET.}
\end{figure}

In a different structure, a p-type NC-FET with a small hysteresis is presented (Figure 6a). A PZT capacitor with an area of 10$\times$10 $\mu m^2$ is connected to the gate of a p-MOSFET (W = 3 $\mu$m, L = 1 $\mu$m). A reduced hysteresis of 200 mV is achieved due to the proper matching of capacitances, regarding equation (6). Figure 5b reports the SS vs. V$_g$ plot which is well below the thermal limit of MOSFETs (down to 20 mV/decade) at 300 K. The internal node measurement confirms a voltage gain greater than 1 while having a peak of 10 V/V (Figure 5c). The polarization vs. voltage plot of the PZT capacitor indicates a clear S-like curve in the positive going branch while it shows a different behavior in the reverse sweep. The ferroelectric performs two separate NC regions, demonstrating a zig-zag polarization characteristic. This mainly happens due to the following reasons; (i) the polycrystalline PZT is showing two main polarization domains despite the applied training procedure \cite{saeidi2018effect} and (ii) a multi-domain ferroelectric can hold one negative capacitance domain at a time \cite{zhu2017negative}. As a result, the manifested polarization characteristic of the multi-domain ferroelectric is different from the S-shaped curve which is expected for a single-domain ferroelectric. Therefore, each domain shows a separate NC region independent of the other one (see Figure 6c), which was also expected from dV$_{int}$/dV$_g$ vs. V$_g$ curve where two individual peaks of the voltage amplification were clearly observed. Figure 5d illustrates the current efficiency enhancement with a maximum value of 400 V$^{-1}$. Figure 5f investigates the impact of the drain-to-source voltage, $|V_{sd}|$, on the input transfer characteristic of the same NC-FET. Besides the common effect of V$_{sd}$ on the level of the drain current, it is evidenced that the NC-FET under lower lateral electric field provides a more effective NC effect and hence, a steeper \textit{off}-to-\textit{on} transition. This is due to the fact that $V_{ds}$ affects the accessible region of the ferroelectric S-curve polarization during the NC-FET operation. An SS of 5 mV/decade is achieved at a $V_{ds}$ of 0.5 V. The \textit{off}-to-\textit{on} transition of NC-FETs with reduced hysteresis (both n- and p-type devices) is not as steep as one of the large hysteresis devices, confirming the proposed theory that a trade-off is needed between the steepness and hysteretic behavior \cite{jo2016negative}. A ferroelectric capacitor that implies a too effective NC results in a large hysteresis together with a sharp transition. Although a super steep switching device is compelling, however, it is not appealing since the reduction of SS is accompanied with a remarkable hysteresis.

\section*{Conclusion}
Energy efficient logic devices are required for the recent advancement of the Internet of Things (IoT) technology. A steep slope field effect transistor with a sub-thermal swing is expected as a key, enabling technology for IoT applications by operating at supply voltages below 0.5 V. Here, it has been shown that the negative capacitance effect can be effectively applied to enhance both digital and analog FoM of advanced CMOS. The measured input transfer characteristics of n- and p-type MOSFETs using PZT as the NC booster shows a steep subthreshold swing down to 10 mV/decade together with an enhanced efficiency factor up to 10$^5$ V$^{-1}$. The on-current over off-current ration is improved and the overdrive is boosted up to 0.45 V. Therefore, the supply voltage can be reduced by 50\%, maintaining the same performance. This is due to the fact that with the aid of a series connected negative capacitor (i.e., with the internal voltage amplification provided by the NC component of the PZT capacitor) the surface potential in MOS devices is increased beyond the applied gate voltage. It has been also demonstrated that the hysteretic behavior of NC-FETs can be tuned considering the proposed stability condition. Both n- and p-type NC-FETs with large (3-4.5 V) and reduced hysteresis (150-200 mV) are implemented, arguing that a trade-off is required between the steepness and hysteretic behavior of an NC-FET. The impact of the drain-to-source electric field on the boosting effect of NC is also demonstrated and discussed, indicating that a lower lateral electric field in the channel results in a steeper \textit{off}-to-\textit{on} transition. It is also experimentally evidenced that a poly-domain ferroelectric cannot have more than one NC domain at a time and shows a zig-zag characteristic.

\begin{acknowledgement}

The authors acknowledge the ERC Advanced Grant Milli-Tech (Grant NO. 695459) for providing the financial support of this research. The authors also greatly appreciate the contributions of Mr. Robin Nigon and Prof. Paul Muralt in the fabrication of the PZT thin film.

\end{acknowledgement}



\begin{thebibliography}{10}
	\expandafter\ifx\csname url\endcsname\relax
	\def\url#1{\texttt{#1}}\fi
	\expandafter\ifx\csname urlprefix\endcsname\relax\def\urlprefix{URL }\fi
	\providecommand{\bibinfo}[2]{#2}
	\providecommand{\eprint}[2][]{\url{#2}}
	
	\bibitem{takagi2008carrier}
	\bibinfo{author}{Takagi, S.} \emph{et~al.}
	\newblock \bibinfo{title}{Carrier-transport-enhanced channel {{CMOS}} for
		improved power consumption and performance}.
	\newblock \emph{\bibinfo{journal}{IEEE Transactions on Electron Devices}}
	\textbf{\bibinfo{volume}{55}}, \bibinfo{pages}{21--39}
	(\bibinfo{year}{2008}).
	
	\bibitem{khandelwal2016circuit}
	\bibinfo{author}{Khandelwal, S.} \emph{et~al.}
	\newblock \bibinfo{title}{Circuit performance analysis of negative capacitance
		{{FinFETs}}}.
	\newblock In \emph{\bibinfo{booktitle}{VLSI Technology, 2016 IEEE Symposium
			on}}, \bibinfo{pages}{1--2} (\bibinfo{organization}{IEEE},
	\bibinfo{year}{2016}).
	
	\bibitem{ionescu2011ultra}
	\bibinfo{author}{Ionescu, A.~M.} \emph{et~al.}
	\newblock \bibinfo{title}{Ultra low power: Emerging devices and their benefits
		for integrated circuits}.
	\newblock In \emph{\bibinfo{booktitle}{Electron Devices Meeting (IEDM), 2011
			IEEE International}}, \bibinfo{pages}{16--1} (\bibinfo{organization}{IEEE},
	\bibinfo{year}{2011}).
	
	\bibitem{salahuddin2008can}
	\bibinfo{author}{Salahuddin, S.} \& \bibinfo{author}{Datta, S.}
	\newblock \bibinfo{title}{Can the subthreshold swing in a classical {{FET}} be
		lowered below 60 mv/decade?}
	\newblock In \emph{\bibinfo{booktitle}{Electron Devices Meeting, 2008. IEDM
			2008. IEEE International}}, \bibinfo{pages}{1--4}
	(\bibinfo{organization}{IEEE}, \bibinfo{year}{2008}).
	
	\bibitem{salahuddin2008use}
	\bibinfo{author}{Salahuddin, S.} \& \bibinfo{author}{Datta, S.}
	\newblock \bibinfo{title}{Use of negative capacitance to provide voltage
		amplification for low power nanoscale devices}.
	\newblock \emph{\bibinfo{journal}{Nano letters}} \textbf{\bibinfo{volume}{8}},
	\bibinfo{pages}{405--410} (\bibinfo{year}{2008}).
	
	\bibitem{zubko2016negative}
	\bibinfo{author}{Zubko, P.} \emph{et~al.}
	\newblock \bibinfo{title}{Negative capacitance in multidomain ferroelectric
		superlattices}.
	\newblock \emph{\bibinfo{journal}{Nature}} \textbf{\bibinfo{volume}{534}},
	\bibinfo{pages}{524--528} (\bibinfo{year}{2016}).
	
	\bibitem{jo2015negative}
	\bibinfo{author}{Jo, J.} \emph{et~al.}
	\newblock \bibinfo{title}{Negative capacitance in organic/ferroelectric
		capacitor to implement steep switching {{MOS}} devices}.
	\newblock \emph{\bibinfo{journal}{Nano letters}} \textbf{\bibinfo{volume}{15}},
	\bibinfo{pages}{4553--4556} (\bibinfo{year}{2015}).
	
	\bibitem{gao2014room}
	\bibinfo{author}{Gao, W.} \emph{et~al.}
	\newblock \bibinfo{title}{Room-temperature negative capacitance in a
		ferroelectric--dielectric superlattice heterostructure}.
	\newblock \emph{\bibinfo{journal}{Nano letters}} \textbf{\bibinfo{volume}{14}},
	\bibinfo{pages}{5814--5819} (\bibinfo{year}{2014}).
	
	\bibitem{ionescu2018negative}
	\bibinfo{author}{Ionescu, A.~M.}
	\newblock \bibinfo{title}{Negative capacitance gives a positive boost}.
	\newblock \emph{\bibinfo{journal}{Nature nanotechnology}}
	\textbf{\bibinfo{volume}{13}}, \bibinfo{pages}{7} (\bibinfo{year}{2018}).
	
	\bibitem{khan2015negative}
	\bibinfo{author}{Khan, A.~I.} \emph{et~al.}
	\newblock \bibinfo{title}{Negative capacitance in a ferroelectric capacitor}.
	\newblock \emph{\bibinfo{journal}{Nature materials}}
	\textbf{\bibinfo{volume}{14}}, \bibinfo{pages}{182} (\bibinfo{year}{2015}).
	
	\bibitem{appleby2014experimental}
	\bibinfo{author}{Appleby, D.~J.} \emph{et~al.}
	\newblock \bibinfo{title}{Experimental observation of negative capacitance in
		ferroelectrics at room temperature}.
	\newblock \emph{\bibinfo{journal}{Nano letters}} \textbf{\bibinfo{volume}{14}},
	\bibinfo{pages}{3864--3868} (\bibinfo{year}{2014}).
	
	\bibitem{yeung2013low}
	\bibinfo{author}{Yeung, C.~W.}, \bibinfo{author}{Khan, A.~I.},
	\bibinfo{author}{Sarker, A.}, \bibinfo{author}{Salahuddin, S.} \&
	\bibinfo{author}{Hu, C.}
	\newblock \bibinfo{title}{Low power negative capacitance fets for future
		quantum-well body technology}.
	\newblock In \emph{\bibinfo{booktitle}{VLSI Technology, Systems, and
			Applications (VLSI-TSA), 2013 International Symposium on}},
	\bibinfo{pages}{1--2} (\bibinfo{organization}{IEEE}, \bibinfo{year}{2013}).
	
	\bibitem{jain2014stability}
	\bibinfo{author}{Jain, A.} \& \bibinfo{author}{Alam, M.~A.}
	\newblock \bibinfo{title}{Stability constraints define the minimum subthreshold
		swing of a negative capacitance field-effect transistor}.
	\newblock \emph{\bibinfo{journal}{IEEE Transactions on Electron Devices}}
	\textbf{\bibinfo{volume}{61}}, \bibinfo{pages}{2235--2242}
	(\bibinfo{year}{2014}).
	
	\bibitem{khan2011ferroelectric}
	\bibinfo{author}{Khan, A.~I.}, \bibinfo{author}{Yeung, C.~W.},
	\bibinfo{author}{Hu, C.} \& \bibinfo{author}{Salahuddin, S.}
	\newblock \bibinfo{title}{Ferroelectric negative capacitance {{MOSFET}}:
		Capacitance tuning \& antiferroelectric operation}.
	\newblock In \emph{\bibinfo{booktitle}{Electron Devices Meeting (IEDM), 2011
			IEEE International}}, \bibinfo{pages}{11--3} (\bibinfo{organization}{IEEE},
	\bibinfo{year}{2011}).
	
	\bibitem{rusu2012analytical}
	\bibinfo{author}{Rusu, A.} \& \bibinfo{author}{Ionescu, A.~M.}
	\newblock \bibinfo{title}{Analytical model for predicting subthreshold slope
		improvement versus negative swing of {{S-shape}} polarization in a
		ferroelectric {{FET}}}.
	\newblock In \emph{\bibinfo{booktitle}{Mixed Design of Integrated Circuits and
			Systems (MIXDES), 2012 Proceedings of the 19th International Conference}},
	\bibinfo{pages}{55--59} (\bibinfo{organization}{IEEE}, \bibinfo{year}{2012}).
	
	\bibitem{saeidi2017negative2}
	\bibinfo{author}{Saeidi, A.} \emph{et~al.}
	\newblock \bibinfo{title}{Negative capacitance field effect transistors;
		capacitance matching and non-hysteretic operation}.
	\newblock In \emph{\bibinfo{booktitle}{Solid-State Device Research Conference
			(ESSDERC), 2017 47th European}}, \bibinfo{pages}{78--81}
	(\bibinfo{organization}{IEEE}, \bibinfo{year}{2017}).
	
	\bibitem{saeidi2016modeling}
	\bibinfo{author}{Saeidi, A.}, \bibinfo{author}{Biswas, A.} \&
	\bibinfo{author}{Ionescu, A.~M.}
	\newblock \bibinfo{title}{Modeling and simulation of low power ferroelectric
		non-volatile memory tunnel field effect transistors using silicon-doped
		hafnium oxide as gate dielectric}.
	\newblock \emph{\bibinfo{journal}{Solid-State Electronics}}
	\textbf{\bibinfo{volume}{124}}, \bibinfo{pages}{16--23}
	(\bibinfo{year}{2016}).
	
	\bibitem{saeidi2017negative}
	\bibinfo{author}{Saeidi, A.} \emph{et~al.}
	\newblock \bibinfo{title}{Negative capacitance as performance booster for
		{{Tunnel FETs}} and {{MOSFETs:}} an experimental study}.
	\newblock \emph{\bibinfo{journal}{IEEE Electron Device Letters}}
	(\bibinfo{year}{2017}).
	
	\bibitem{saeidi2016double}
	\bibinfo{author}{Saeidi, A.}, \bibinfo{author}{Jazaeri, F.},
	\bibinfo{author}{Stolichnov, I.} \& \bibinfo{author}{Ionescu, A.~M.}
	\newblock \bibinfo{title}{Double-gate negative-capacitance {{MOSFET}} with
		{{PZT}} gate-stack on ultra thin body {{SOI:}} an experimentally calibrated
		simulation study of device performance}.
	\newblock \emph{\bibinfo{journal}{IEEE Transactions on Electron Devices}}
	\textbf{\bibinfo{volume}{63}}, \bibinfo{pages}{4678--4684}
	(\bibinfo{year}{2016}).
	
	\bibitem{karda2016switching}
	\bibinfo{author}{Karda, K.}, \bibinfo{author}{Mouli, C.} \&
	\bibinfo{author}{Alam, M.}
	\newblock \bibinfo{title}{Switching dynamics and hot atom damage in landau
		switches}.
	\newblock \emph{\bibinfo{journal}{IEEE Electron Device Letters}}
	\textbf{\bibinfo{volume}{37}}, \bibinfo{pages}{801--804}
	(\bibinfo{year}{2016}).
	
	\bibitem{li2015sub}
	\bibinfo{author}{Li, K.-S.} \emph{et~al.}
	\newblock \bibinfo{title}{Sub-60mv-swing negative-capacitance {{FinFET}}
		without hysteresis}.
	\newblock In \emph{\bibinfo{booktitle}{Electron Devices Meeting (IEDM), 2015
			IEEE International}}, \bibinfo{pages}{22--6} (\bibinfo{organization}{IEEE},
	\bibinfo{year}{2015}).
	
	\bibitem{jimenez2010analytic}
	\bibinfo{author}{Jimenez, D.}, \bibinfo{author}{Miranda, E.} \&
	\bibinfo{author}{Godoy, A.}
	\newblock \bibinfo{title}{Analytic model for the surface potential and drain
		current in negative capacitance field-effect transistors}.
	\newblock \emph{\bibinfo{journal}{IEEE Transactions on Electron Devices}}
	\textbf{\bibinfo{volume}{57}}, \bibinfo{pages}{2405--2409}
	(\bibinfo{year}{2010}).
	
	\bibitem{aziz2016physics}
	\bibinfo{author}{Aziz, A.}, \bibinfo{author}{Ghosh, S.},
	\bibinfo{author}{Datta, S.} \& \bibinfo{author}{Gupta, S.~K.}
	\newblock \bibinfo{title}{Physics-based circuit-compatible spice model for
		ferroelectric transistors}.
	\newblock \emph{\bibinfo{journal}{IEEE Electron Device Letters}}
	\textbf{\bibinfo{volume}{37}}, \bibinfo{pages}{805--808}
	(\bibinfo{year}{2016}).
	
	\bibitem{hoffmann2017modeling}
	\bibinfo{author}{Hoffmann, M.}, \bibinfo{author}{Pe{\v{s}}i{\'c}, M.},
	\bibinfo{author}{Slesazeck, S.}, \bibinfo{author}{Schroeder, U.} \&
	\bibinfo{author}{Mikolajick, T.}
	\newblock \bibinfo{title}{Modeling and design considerations for negative
		capacitance field-effect transistors}.
	\newblock In \emph{\bibinfo{booktitle}{Ultimate Integration on Silicon
			(EUROSOI-ULIS), 2017 Joint International EUROSOI Workshop and International
			Conference on}}, \bibinfo{pages}{1--4} (\bibinfo{organization}{IEEE},
	\bibinfo{year}{2017}).
	
	\bibitem{kidoh1991ferroelectric}
	\bibinfo{author}{Kidoh, H.}, \bibinfo{author}{Ogawa, T.},
	\bibinfo{author}{Morimoto, A.} \& \bibinfo{author}{Shimizu, T.}
	\newblock \bibinfo{title}{Ferroelectric properties of lead-zirconate-titanate
		films prepared by laser ablation}.
	\newblock \emph{\bibinfo{journal}{Applied physics letters}}
	\textbf{\bibinfo{volume}{58}}, \bibinfo{pages}{2910--2912}
	(\bibinfo{year}{1991}).
	
	\bibitem{nakamura1994preparation}
	\bibinfo{author}{Nakamura, T.}, \bibinfo{author}{Nakao, Y.},
	\bibinfo{author}{Kamisawa, A.} \& \bibinfo{author}{Takasu, H.}
	\newblock \bibinfo{title}{Preparation of {{Pb (Zr, Ti) O$_3$}} thin films on
		electrodes including {{IrO$_2$}}}.
	\newblock \emph{\bibinfo{journal}{Applied physics letters}}
	\textbf{\bibinfo{volume}{65}}, \bibinfo{pages}{1522--1524}
	(\bibinfo{year}{1994}).
	
	\bibitem{khan2016negative}
	\bibinfo{author}{Khan, A.~I.}, \bibinfo{author}{Radhakrishna, U.},
	\bibinfo{author}{Chatterjee, K.}, \bibinfo{author}{Salahuddin, S.} \&
	\bibinfo{author}{Antoniadis, D.~A.}
	\newblock \bibinfo{title}{Negative capacitance behavior in a leaky
		ferroelectric}.
	\newblock \emph{\bibinfo{journal}{IEEE Transactions on Electron Devices}}
	\textbf{\bibinfo{volume}{63}}, \bibinfo{pages}{4416--4422}
	(\bibinfo{year}{2016}).
	
	\bibitem{lin2016effects}
	\bibinfo{author}{Lin, C.-I.}, \bibinfo{author}{Khan, A.~I.},
	\bibinfo{author}{Salahuddin, S.} \& \bibinfo{author}{Hu, C.}
	\newblock \bibinfo{title}{Effects of the variation of ferroelectric properties
		on negative capacitance {{FET}} characteristics}.
	\newblock \emph{\bibinfo{journal}{IEEE Transactions on Electron Devices}}
	\textbf{\bibinfo{volume}{63}}, \bibinfo{pages}{2197--2199}
	(\bibinfo{year}{2016}).
	
	\bibitem{saeidi2018effect}
	\bibinfo{author}{Saeidi, A.} \emph{et~al.}
	\newblock \bibinfo{title}{Effect of hysteretic and non-hysteretic negative
		capacitance on tunnel {{FETs DC}} performance}.
	\newblock \bibinfo{type}{Tech. Rep.}, \bibinfo{institution}{Institute of
		Physics} (\bibinfo{year}{2018}).
	
	\bibitem{khan2016negative2}
	\bibinfo{author}{Khan, A.~I.} \emph{et~al.}
	\newblock \bibinfo{title}{Negative capacitance in short-channel finfets
		externally connected to an epitaxial ferroelectric capacitor}.
	\newblock \emph{\bibinfo{journal}{IEEE Electron Device Letters}}
	\textbf{\bibinfo{volume}{37}}, \bibinfo{pages}{111--114}
	(\bibinfo{year}{2016}).
	
	\bibitem{jo2016negative}
	\bibinfo{author}{Jo, J.} \& \bibinfo{author}{Shin, C.}
	\newblock \bibinfo{title}{Negative capacitance field effect transistor with
		hysteresis-free sub-60-mv/decade switching}.
	\newblock \emph{\bibinfo{journal}{IEEE Electron Device Letters}}
	\textbf{\bibinfo{volume}{37}}, \bibinfo{pages}{245--248}
	(\bibinfo{year}{2016}).
	
	\bibitem{rusu2016condition}
	\bibinfo{author}{Rusu, A.}, \bibinfo{author}{Saeidi, A.} \&
	\bibinfo{author}{Ionescu, A.~M.}
	\newblock \bibinfo{title}{Condition for the negative capacitance effect in
		metal--ferroelectric--insulator--semiconductor devices}.
	\newblock \emph{\bibinfo{journal}{Nanotechnology}}
	\textbf{\bibinfo{volume}{27}}, \bibinfo{pages}{115201}
	(\bibinfo{year}{2016}).
	
	\bibitem{zhu2017negative}
	\bibinfo{author}{Zhu, Z.} \emph{et~al.}
	\newblock \bibinfo{title}{Negative-capacitance characteristics in a
		steady-state ferroelectric capacitor made of parallel domains}.
	\newblock \emph{\bibinfo{journal}{IEEE Electron Device Letters}}
	\textbf{\bibinfo{volume}{38}}, \bibinfo{pages}{1176--1179}
	(\bibinfo{year}{2017}).
	
\end{thebibliography}
\end{document}